\documentclass[]{interact}

\usepackage{epstopdf}
\usepackage[caption=false]{subfig}
\usepackage[natbibapa,nodoi]{apacite}
\bibpunct[, ]{(}{)}{;}{a}{,}{,}

\usepackage{graphicx}
\usepackage{float}
\usepackage{longtable}
\usepackage{tabularx}
\usepackage{color}
\usepackage{multirow}
\usepackage{array}
\usepackage{gensymb}
\usepackage{geometry}
\usepackage{pdflscape}
\usepackage[colorlinks=true, allcolors=blue]{hyperref}
\usepackage{adjustbox}
\usepackage{xcolor}

\theoremstyle{plain}

\theoremstyle{definition}

\theoremstyle{remark}

\begin{document}

\articletype{Research Article}

\title{Traffic Co-Simulation Framework Empowered by Infrastructure Camera Sensing and Reinforcement Learning}


\author{ \name{Talha Azfar\textsuperscript{a}\thanks{CONTACT Ruimin Ke. Email: ker@rpi.edu}, Kaicong Huang\textsuperscript{a}, Andrew Tracy\textsuperscript{b}, Sandra Misiewicz\textsuperscript{b}, Chenxi Liu\textsuperscript{c}, and Ruimin Ke\textsuperscript{a}}
\affil{\textsuperscript{a} Department of Civil and Environmental Engineering, Rensselaer Polytechnic Institute, Troy, NY 12180, USA \\
\textsuperscript{b} Capital Region Transportation Council, Albany, NY 12205, US \\
\textsuperscript{c} Department of Civil and Environmental Engineering, University of Utah, Salt Lake City, UT 84112, USA }}

\maketitle

\begin{abstract}
Traffic simulations are commonly used to optimize urban traffic flow, with reinforcement learning (RL) showing promising potential for automated traffic signal control, particularly in intelligent transportation systems involving connected automated vehicles. Multi-agent reinforcement learning (MARL) is particularly effective for learning control strategies for traffic lights in a network using iterative simulations. However, existing methods often assume perfect vehicle detection, which overlooks real-world limitations related to infrastructure availability and sensor reliability. 
This study proposes a co-simulation framework integrating CARLA and SUMO, which combines high-fidelity 3D modeling with large-scale traffic flow simulation. Cameras mounted on traffic light poles within the CARLA environment use a YOLO-based computer vision system to detect and count vehicles, providing real-time traffic data as input for adaptive signal control in SUMO. MARL agents trained with four different reward structures leverage this visual feedback to optimize signal timings and improve network-wide traffic flow. 
Experiments in a multi-intersection test-bed demonstrate the effectiveness of the proposed MARL approach in enhancing traffic conditions using real-time camera based detection. The framework also evaluates the robustness of MARL under faulty or sparse sensing and compares the performance of YOLOv5 and YOLOv8 for vehicle detection. Results show that while better accuracy improves performance, MARL agents can still achieve significant improvements with imperfect detection, demonstrating scalability and adaptability for real-world scenarios.
\end{abstract}


\section{Introduction}
Traffic congestion is a significant issue in urban areas leading to increased travel times, fuel consumption, and environmental pollution, causing considerable economic impact \citep{goodwin2004economic, luan2022traffic, qiu2020carbon}. Effective traffic signal control (TSC) can alleviate these problems by optimizing the flow of vehicles through intersections. Traditional TSC methods often rely on fixed-time schedules or simple rule-based systems, which may not adapt well to dynamic and complex traffic conditions \citep{mousavi2017traffic}. In the Capital Region of New York State, most locally-owned (non-DOT) signals are pre-timed, and all DOT signals are actuated, but none are coordinated for network-wide optimization except a small number of local signals. As a result, there is growing interest in applying advanced techniques, such as reinforcement learning (RL), to develop adaptive and efficient traffic signal control systems.

Evaluating traffic control strategies in simulation before real-world deployment is essential for intelligent transportation systems (ITS) \citep{feng2023dense}. Among the available simulation tools, Simulation of Urban MObility (SUMO) \citep{krajzewicz2002sumo} stands out for its open-source availability, ease of use, and effectiveness \citep{li2024chatsumo}. While this is a sufficient and widely used environment for traffic signal control experimentation, the lack of realistic synthetic sensor data limits the applicability to real-world situations. Despite the widespread use of video cameras due to their low cost and ability to capture rich data \citep{ke2020real}, previous studies have yet to fully integrate them into SUMO-based traffic control systems. To bridge this gap, co-simulation frameworks that combine multiple simulators are gaining popularity as a means to create more realistic and comprehensive evaluation environments that include virtual cameras \citep{he2024holistic}. However, further work is needed to transform these cameras into smart sensors that can interact with microsimulation control algorithms in a synchronized way. This integration would create a virtual test-bed with significant potential to simulate surveillance camera-based ITS control strategies.

In this study, we propose and evaluate such a promising test-bed by exploring the use of multi-agent reinforcement learning (MARL) for traffic signal optimization and using camera-based detection for testing. 
In order to examine the effectiveness of the co-simulation framework, we develop and evaluate MARL agents in this realistic test-bed that integrates the CARLA simulator \citep{Dosovitskiy17} and the SUMO traffic simulator. CARLA is an open-source urban driving simulator that provides high-fidelity 3D environments for vehicle dynamics and sensor simulations, while SUMO is an open-source, highly portable, microscopic and continuous multi-modal traffic simulation package designed to handle large road networks \citep{ke2023digital}. 

MARL involves training multiple RL agents, each controlling a traffic signal, to collectively improve overall travel time in the road network. This approach allows each traffic signal to adapt its timings based on real-time traffic conditions, leading to more efficient traffic management \citep{wang2020large}. There are numerous ways to train these agents in terms of architecture and reward functions. However, the challenges of real-world applicability stemming from faulty or sparse sensing, communication delays, and computational bottlenecks can significantly impact the effectiveness of any control system \citep{azfar2024deep}, yet these issues are rarely addressed in simulation studies. This consideration is directly relevant to the application of connected automated vehicles (CAV), which depend on infrastructure-based sensing and communication. Demonstrating traffic control strategies that are resilient to noisy detection is therefore essential for supporting CAV deployment in practice.

\begin{figure*}[!p]
    \centering
    \includegraphics[width=0.85\linewidth]{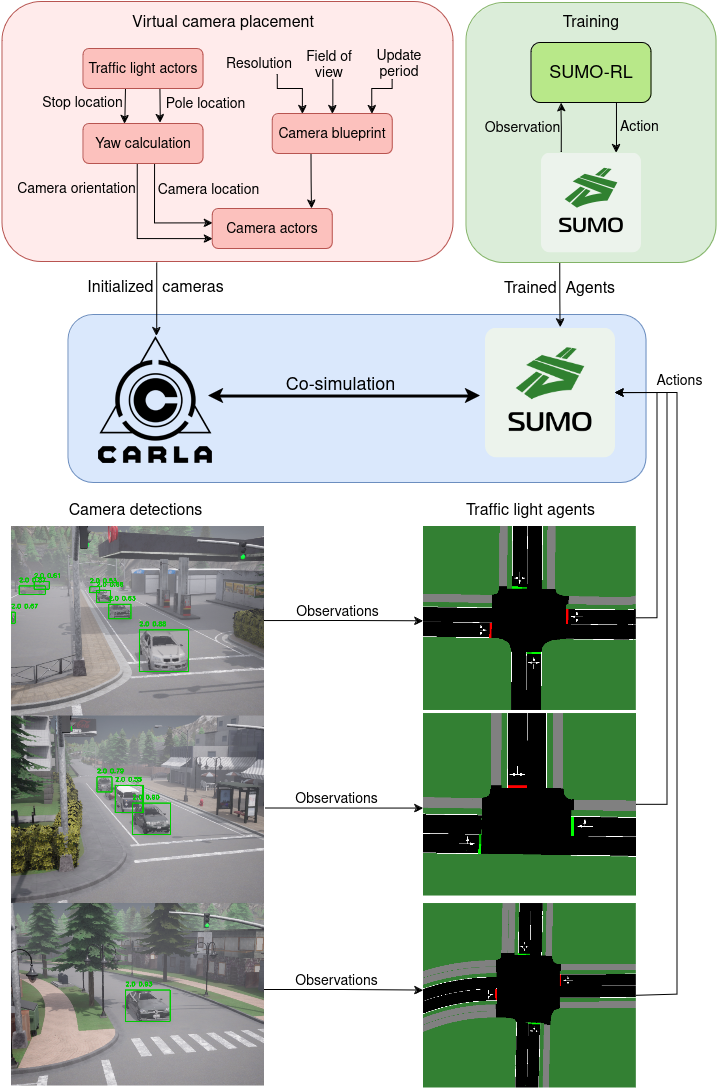}
    \caption{Overview of the framework}
    \label{fig:framework}
\end{figure*}

In our framework, CARLA is used to provide a realistic urban environment and detailed vehicle dynamics, while SUMO handles traffic flow and signal control. Cameras are deployed on traffic light poles within the CARLA environment to monitor traffic at intersections. In the Capital Region of New York State, most signals are on span wire, so cameras need to go on the poles.  These cameras feed a convolutional neural network based computer vision system to detect and count vehicles, providing real-time data on traffic conditions which is used by the MARL agents to control the next phase of the traffic light. Figure \ref{fig:framework} showcases the framework for the system. The agents have been trained using a Q-learning algorithm to optimize signal timings based on the observed traffic conditions. The goal is to minimize network-wide vehicle waiting times and improve overall traffic flow. This is the most common measure of effectiveness (MoE), while other MoEs like number of arrivals on red could also be used because it correlates with quality of traveler experience and vehicle emissions.

The co-simulation approach offers significant advantages over traditional single-simulator methods. By leveraging the strengths of both CARLA and SUMO, this framework enables the evaluation of MARL-based TSC in a highly realistic and dynamic environment. This setup not only captures the complexity of real-world traffic conditions but also allows for the testing of intelligent sensor integration, making it a powerful tool for advancing TSC research. Our goal is to demonstrate the potential of this co-simulation framework in minimizing network-wide vehicle waiting times and improving overall traffic flow using infrastructure camera input, thus paving the way for more effective and deployable ITS solutions.

This paper presents the methodology and results of our co-simulation study, demonstrating the potential of MARL for adaptive TSC using computer vision based detection that is often imperfect. We discuss the relevant literature followed by the setup of the CARLA-SUMO co-simulation environment, the implementation of the vision system for vehicle detection, and the training and evaluation of the MARL agents. Our results show that MARL-based TSC can significantly improve traffic flow compared to traditional fixed-time or actuated signal control methods while using inputs from infrastructure camera-based vehicle detection. We also discuss the errors related to detection and its effects on the system performance. In summary, the primary contributions of the paper are as follows:

\begin{itemize}
    \item The study proposes a co-simulation framework that utilizes CARLA for high-fidelity 3D environments and SUMO for traffic flow and signal control. The integration allows for the use of computer vision-based vehicle detection in CARLA as input for large-scale traffic signal control in SUMO.
    
    \item The framework programmatically accesses traffic light pole locations in the CARLA environment, spawns virtual cameras at these locations, and uses coordinate matching between CARLA and SUMO to identify the stopping line of the lane each light serves. It then automatically calculates the camera's yaw angle to ensure it is properly aimed. These cameras use a YOLO-based computer vision system to detect and count vehicles, providing real-time data on traffic conditions in the virtual test-bed.
    
    \item The study employs MARL with four different well known rewards to train multiple agents for adaptive traffic signal control, optimizing the timings based on feedback of real-time traffic conditions to improve network-wide traffic flow. Experiments on the co-simulation test-bed show that the proposed MARL significantly improves traffic signal control using camera-based detection.
    
    \item The MARL agents are in scenarios with potentially faulty or sparse sensing, highlighting the importance of robustness in real-world applications. The paper examines performance differences between YOLOv5 and the newer YOLOv8, and how the detection errors affect the agents trained using different rewards, finding that while increased accuracy improves performance, even imperfect detection can yield positive results with the right MARL training.
\end{itemize}

\section{Literature Review}

There are numerous traffic simulation software available for different purposes including traffic flow modeling, simulating connected autonomous vehicles, vehicle dynamics and emissions modeling, and traffic signal control \citep{nguyen2021overview}. Among the most widely used traffic simulators are AIMSUN \citep{casas2010traffic}, VISSIM \citep{fellendorf2010microscopic}, and MATSim \citep{balmer2009matsim}, each offering unique features and capabilities to suit different research needs \citep{diallo2021comparative}. However, the focus of this study is on SUMO \citep{krajzewicz2002sumo}, a highly regarded open-source traffic simulator. The open-source nature of SUMO and the Traffic Control Interface (TraCI) enable finely customizable real-time programmable interaction with the simulation. It has a rich history of use in intelligent traffic signal control for example, \cite{AzevedoARR16} created a toolchain to access and control SUMO traffic lights, enabling machine learning via external programming, and \cite{guo2019reinforcement} developed a Q-network for a single intersection in SUMO. 

While automated traffic signal control has been around for decades, most notably the Sydney coordinated adaptive traffic system (SCATS) \citep{sims1980sydney} and  split, cycle, and offset optimisation technique (SCOOT) \citep{hunt1981scoot}. These methods utilized real-time traffic flow data at each intersection to manage the timing of signal phases. The SCATS system uses vehicle counts at one intersection whereas SCOOT relies on upstream detectors for more detailed flow information. Both systems are capable of coordination with neighboring intersections by communicating their state information. However, the systems operate in a local context, their behavior is limited by the preprogrammed algorithm, and they are sensitive to the position and reliability of the vehicle sensors \citep{reza2023citywide}.

Machine learning strategies, in particular reinforcement learning (RL) has emerged as an effective strategy for automated traffic control \citep{zhu2020safe}. Coordinated deep Q-network RL was used with transfer planning for fast and scalable learning, but the training was found to occasionally lack in stability \citep{van2016coordinated}. 
Reinforced Signal Control (RESCO) \citep{ault2021} provided a benchmark test-bed for RL based TSC inspired by real world cities and calibrated demand. Their extensive experimentation showed that decentralized control algorithms are more robust as compared to coordinated control methods, which are hampered by stricter sensing requirements and do not generalize well to all traffic conditions. A scalable multi-agent RL (MARL) algorithm for TSC was developed in \cite{chu2019multi} which features independent advantage actor critic RL agents, each of which is kept updated about its neighbor's policy. It was tested on a synthetic traffic grid and Monaco city traffic network, with the salient limitations being a lack of robustness to noisy or delayed sensor measurements, high memory consumption of the agent models, and communication latency between neighboring intersections. \cite{xiao2022cold} introduced a staged dueling double deep Q-network integrated with model based control to avoid the cold start problem for RL. A strategy to simplify the optimization process in \cite{xu2020network} identifies the critical nodes using trajectory based CRRank analysis and only applies LSTM based RL for network-wide control. Asynchronous n-step Q-learning was used to train an adaptive TSC reducing delays compared to actuated control in \cite{genders2019asynchronous}, but increased delays for left turns indicated the need to fine-tune the reward function.

Regarding the interpretability of intelligent traffic control agents, it is demonstrated in \cite{ault20} that a regulatable control policy is one in which changes to the signal assignment can be understood as a direct consequence of changes in the state of the traffic. This is important for safety and regulatory considerations and is comparable to the actuated signal technology currently deployed based on government standards \citep{bonneson2011traffic}.

A traffic simulator such as SUMO can perform its role effectively but is limited in its ability to provide synthetic sensor data beyond loop detectors. Additional simulation tools are required for problems that require real-world considerations such as demand prediction, advanced sensing, visual accuracy, vehicle mechanics, etc. These simulations must be synchronized for consistency and validity, and this process is known as co-simulation. MATSim was integrated with external simulators for fleet demand simulation in \cite{yang2024co} to showcase ride-pooling services with agent based simulation. 

CARLA is an open source simulator for autonomous driving built atop the Unreal 4 game engine \citep{Dosovitskiy17}. It features high quality 3D graphics, a flexible API, and integration with SUMO. CARLA-SUMO co-simulation is extensively used in many applications including the study of connected and autonomous vehicles \citep{stepanyants2023survey}, real-time sensor fusion \citep{cantas2021customized}, computer vision based cooperative perception \citep{aoki2020cooperative}, and  transportation digital twins \citep{azfar2022efficient}. The co-simulation has also been used in conjunction with image processing and object detection to estimate traffic flow statistics such as volume and trajectories from real world video in order to generate simulation scenarios based on real-world conditions \citep{shi2022integrated}. 

Recently, co-simulation of SUMO and MetaDrive was used to train different RL algorithms for TSC at a single intersection using the raw pixel values of the image as observation input to the learning mechanism \citep{he2024holistic}. This incurs a notable performance cost of training on high dimensional data but allows setting up different visual conditions, thus reducing the simulation-reality gap. Frameworks like MetaDrive+SUMO, while valuable, lack the visual fidelity required to test the robustness of vision algorithms against realistic environmental conditions. Our use of CARLA provides high-quality 3D graphics that allow for testing under varied weather and lighting, which is crucial for evaluating performance in real-world scenarios.

Real-time video from 14 traffic cameras was used for vehicle detection and tracking using transfer learning, and the extracted information was incorporated into a simulation for TSC optimization \citep{ZHOU2021102775}. The optimization was performed in simulation after analyzing previous 15 minutes of video and claims to reduce 14.2\% of average vehicle delays. In another example of computer vision based signal timing control, vehicles were detected from video and represented in temporal clusters which were used to estimate road occupancy instead of counting \citep{kumaran2019computer}. Phase durations were determined from Gaussian regression, and while the inference process was fast enough for real-time execution, the framework was only valid for a single intersection. Similarly in \cite{gorodokin2021optimization}, YOLOv4 was used to count vehicles and Greenshield's model was used to adjust the actuation of the green time for a single intersection. The YOLO family of convolutional neural network based models are some of the most popular object detection models, designed to be fast and accurate with pretrained checkpoints available in various sizes \citep{yolov5}. Aerial images from the VISSIM simulator were used to train RL-based signal control for a single intersection in \cite{jeon2018artificial}, providing an inconsistent but significant reduction in delays. 

Other approaches incorporate real-time information through probe vehicles \citep{lian2021adaptive} however, with limited penetration rates, their usefulness is largely restricted to congested conditions, while under normal flow they serve mainly as proxies for average speed. Multi-agent optimization based on dynamic programming and optimal control demonstrate superior performance but assume perfect communication and observations \citep{xu2019optimizing}. The use of cellular data to estimate traffic flows is investigated in \cite{yang2016performance} by including a realistic simulation of wireless signals in a traffic environment. However, in practice, traffic cameras are already widely deployed for monitoring and incident detection, and have been shown to be highly cost-effective, returning up to \$12 in benefits for every dollar spent \citep{fries2007accelerated}.

Reinforcement learning for TSC is generally performed using ground truth observations from the simulator itself, and the testing is also done in the same environment with varying traffic demands. 
A brief selection of papers involving co-simulation are compared in Table \ref{Lit-Table} with applications such as connected autonomous vehicles, traffic modeling, and traffic signal control.

\begin{table*}[!htbp] 
\renewcommand{\arraystretch}{1.5} 
\centering 
      \small
\caption{Existing literature related to traffic co-simulation }
\label{Lit-Table}
\adjustbox{rotate=90,center}{
\begin{tabular}{p{8.5em} p{13em} p{13em} p{3.5em} p{8em} p{3.5em} p{3.5em}}
\multicolumn{1}{c}{Paper}                        & Task                           & Key Method    & Roadside Sensing    &  \hspace{0.05cm}  Scale     & Feedback Loop    & Visual Realism       \\ \hline \hline
\cite{osinski2020simulation} & RL for autonomous driving        & Image segmentation + CNN       &  No               & Single vehicle  &    \textbf{Yes}  &  \textbf{Yes}  \\ 
 \cite{cantas2021customized}  & Collaborative perception for CAVs & LiDAR 3D detection + Sensor fusion   & No            & Multiple vehicles  & \textbf{Yes}   &  \textbf{Yes} \\ 
 \cite{aoki2020cooperative}      & Cooperative perception and V2X communication  & Object classification + Deep-Q RL    & No & Multiple vehicles &  No   &  \textbf{Yes}  \\ 
 \cite{azfar2023incorporating}   & Incorporating vehicle detection in Digital twin     & Edge-based vehicle detection      & \textbf{Yes} & Single roundabout  & No  &  Real video  \\ 
 \cite{shi2022integrated}     & Data-driven test-bed for traffic scenes   & Vehicle detection and trajectory extraction  & \textbf{Yes} &  Highway segment or intersection  & No & Real video  \\ 
  \cite{ZHOU2021102775}   & CV based traffic modeling and adaptive signal schemes   & Transfer learning for vehicle detection and tracking  &  \textbf{Yes}   & Road network & No & No  \\  
 \cite{kumaran2019computer}   & CV based traffic signal optimization     & Optical flow and clustering + Gaussian regression    &  \textbf{Yes}  & Single intersection & No &  Real video \\  
 \cite{he2024holistic}              & CV based isolated TSC   & End-to-end deep RL from images   & \textbf{Yes} &  Single intersection  & \textbf{Yes}    &   No    \\  
\cite{zhao2022co}  &  Modeling and evaluating CAV and human behavior  & Mixed traffic modeling and game theory  & No &  Highway on-ramp  & \textbf{Yes} &  \textbf{Yes}  \\
 \cite{insuasty2019control}  &  Control strategies for urban traffic with co-simulation  &  Particle swarm optimization + Replicator dynamics  & No & Two intersections & \textbf{Yes}  &  No \\
 \cite{zhu2024evaluating} & Mixed traffic stability analysis in co-simulation & Mathematical modeling with V2V and car following & No & Multiple vehicles & No &  No \\
 \cite{jeon2018artificial} & Traffic signal control from VISSIM aerial images & Deep CNN for Q function approximation &  No & Single intersection & No & Real video  \\ 
Ours                                             & Test-bed for multi-agent traffic signal optimization using infrastructure camera input      & Vehicle detection + MARL    & \textbf{Yes}   &  \textbf{Multiple intersections } & \textbf{Yes}  &  \textbf{Yes}  \\ \hline
\end{tabular}
}
\end{table*}

To the best of our knowledge, this is the first study to explicitly model and evaluate a network of infrastructure-mounted cameras for multi-agent reinforcement learning (MARL)-based traffic signal control. No existing simulation framework currently supports MARL-based signal control using realistically rendered, camera-based detection. To address this gap, we propose and demonstrate a visually realistic co-simulation environment that integrates video detection in CARLA with signal control in SUMO. The core of our evaluation is testing MARL robustness to imperfect sensing from realistic, ground-level cameras. CARLA runs on Unreal Engine and boasts realistic dynamic lighting, weather, and reflections. This cannot be replicated in low-fidelity simulators like MetaDrive, which trades off realistic visual rendering for performance. Other studies that utilize aerial imagery with VISSIM simplify the problem by adopting a bird’s-eye view, thereby bypassing the complexities associated with modeling ground-level camera infrastructure. Our framework is uniquely positioned to close the `simulation-reality gap' for this specific problem.

\section{Methodology}

\subsection{Overview}
The aim of the study was to create a co-simulation framework that supports the optimization of any number of traffic lights in a road network for real-time adaptive traffic management based on computer vision input. It was implemented using CARLA and SUMO, with the traffic signal optimization being performed by multiagent Q-learning. SUMO serves as a reliable and highly customizable traffic microsimulator capable of generating training environment and data for reinforcement learning, while CARLA presents a realistic 3D rendering of the urban landscape complete with a diverse array of vehicles that are being controlled by SUMO. 

The process starts with using a SUMO road network with traffic lights as Q-learning agents utilizing the \verb|sumo_rl| Python library \citep{sumorl} which converts SUMO networks to RL environments based on the stable baselines format, allowing the use of standard RL algorithms that the library offers. Each traffic light is an independent agent taking the current phase, occupancy of incoming lanes, and queue length as the state. Each agent calculates its reward from its incoming lanes based on one of four reward functions tested. The trained list of agents is saved to a Python pickle file to be used in the co-simulation.

The CARLA simulation was configured to run the same map as the one used for training, and cameras were spawned directly above each traffic light pole. The coordinates of the stopping point of the lane being served by each light were used to point the yaw angle of the camera, along with a 10 degrees downward pitch. Every camera captures its image once per second and stores it to memory. YOLOv5 is used to detect vehicles in the oncoming lane by selecting a region around the center of the camera view so that other irrelevant vehicles are not counted. The object detection is performed for each intersection periodically, and the observation is fed into the corresponding agent from the list in the loaded pickle file. The resulting action of the agent is to choose the next phase for the intersection which is then passed to the SUMO simulation. The SUMO-CARLA bridge synchronizes both simulations to ensure accurate visual representation of the scenario.

The outlined  approach represents a virtual test bed for the real world deployment of a computer vision based, distributed and automated, real-time traffic signal control system. The performance was evaluated by comparing the average speed, wait times, and queue lengths before and after the implementation of the multiagent Q-learning approach. 

\subsection{Vehicle Detection and Aggregation using Virtual Infrastructure Cameras}
The CARLA simulation environment consists of 3D mesh objects of the environment, a road network, and so-called actors representing dynamic agents such as vehicles, traffic lights, and sensors. These actors may be created as part of the map or spawned during the simulation. In this work, we use the existing traffic lights as the location to mount our virtual cameras. The CARLA python library is used to access the traffic light objects and obtain their 3D location, while the RGB camera blueprint is queried from the blueprint library and initialized with a resolution of $640 \times 480$ pixels and a field of view of $45 \degree$. As real-time tracking was not required, and to reduce the computational burden, each camera update was set to a period of 1 second. 

The CARLA-SUMO co-simulation Python file contains a bridge helper object which is used to get the corresponding locations between the two environments since they each operate with their separate coordinate systems. This function is used to match the CARLA traffic lights to SUMO junctions and to obtain the location of the corresponding stopping line for each traffic light. The yaw angle for the camera is then calculated using the camera position and the stopping line position by taking the inverse tangent of the ratio of the difference in $y$ to the difference in $x$ coordinates. Pitch angle is set to $-10 \degree$ and roll angle is kept zero, ensuring that the concerning lane for each camera appears in the middle of the image frame. Using this method cameras can be automatically spawned on all traffic lights while ensuring correct placement and orientation.

Upon every capture the camera image is saved to memory and used as input to a vehicle detection algorithm. For this purpose we used YOLOv5 as it is simple to implement, executes fast, and provides accurate results. The particular model used in this implementation was the YOLOv5m version, which has 21.2 million parameters. The detections are performed on a $240 \times 480$ pixel region around the center of the image, and are filtered to only include cars, trucks, buses, and motorcycles. Since each camera is associated with one traffic light pole, all the cameras from an intersection can potentially give complete information about the vehicles queued in the approaching lanes. This information is used to build the observation state of the traffic light control agent that has been trained using reinforcement learning.

\subsection{Reinforcement Learning Agent Training}
The SUMO road network of the map used in the co-simulation is used as the environment for reinforcement learning for traffic signal optimization. Each traffic signal is an independent reinforcement learning agent using Q-learning to learn a policy that maximizes the total reward over time. The main reason to choose Q learning, in addition to its simplicity, is that after training, the Q-table can be inspected and there is no chance of unexplained behavior in implementation. 
The reward for each agent may based on any relevant information in the simulation, for example, the difference in accumulated waiting time per lane between time steps. A decreasing waiting time results in higher reward. Each agent contains a Q-table that is iteratively updated during training by first choosing an action using the $\varepsilon$-greedy policy that chooses a random action with a probability of $\varepsilon$, or the action with the highest Q-value from the table for the current state. The Q-table is a matrix where each row corresponds to a state in the environment, and each column corresponds to an action that the agent can take in that state. Over time, the value of $\varepsilon$ is decreased so that the learned Q-table is exploited more often. Once the action is performed, the next state and reward are observed and used to update the Q-table using equation 1. 
\begin{equation}
Q(s, a) \leftarrow Q(s, a)+\alpha\left(r + \gamma \max _{a^{\prime}} Q\left(s^{\prime}, a^{\prime}\right)-Q(s, a)\right)
\end{equation}
where $\alpha$ is the learning rate, $\gamma$ is a discount factor that determines the importance of future rewards, $r$ is the reward after taking action $a$ from state $s$, and $s'$ is the new state after the action. $\max_{a'} Q(s', a')$ refers to the maximum Q-value for the next state over all possible actions $a'$. 

The learning steps are repeated for multiple episodes, where one episode is the entirety of the simulation when all vehicles have completed their routes. 

\textbf{Agent: } Let agent $i \in \{1, ...,N \}$ control a signalized intersection. Each traffic light is a reinforcement learning agent with its own state space, action space, learning rate, discount factor, and exploration strategy. While these variables can be made different for experimentation, in this work we initialize all agents with the same parameters. The action and state spaces, however, depend on the incoming lanes and number of phases.

\textbf{State space:} The state of the $i$th agent at time $t$ can be represented as $$s_i^t=\left(p_i^t, m_i^t, \mathbf{d}_i^t, \mathbf{q}_i^t\right)$$
It is an observation consisting of the current phase index $p^t_i \in \{0,1,...P-1\}$, a boolean variable $m^i_t \in \{0,1\}$ indicating whether minimum green time has elapsed, occupancy vector $\mathbf{d}^t_i \in \mathbb{N}^{L_i}$ of each incoming lane $L$ at intersection $i$, and queue vector $\mathbf{q}^t_i \in \mathbb{N}^{L_i}$ of each incoming lane. The occupancy is defined as the number of vehicles approaching the intersection from each incoming lane, while the queue is the number of stopped vehicles in each incoming lane. As there is no assumption of communication between agents, any information external to an agent is not available to it for decision making. 

\textbf{Action space: } The action is the integer index of the green phase to be activated at the next time step. Every agent can have a different action space corresponding to the number of green phases it has. Each agent $i$ selects an action $a^t_i \in \mathcal{A}_i$, where the action space is $\mathcal{A}_i=\left\{0,1, \ldots, G_i-1\right\}$. 
Here, $G_i$ is the number of valid green phases for intersection $i$. Each element corresponds to a unique phase (e.g., protected left turn, through traffic), as defined in the SUMO network.
The environment controlling the simulation ensures that the appropriate yellow phase is inserted between different phases, and there is no implementation of an all red phase. For intersections with complex movements, each legally permissible combination of movements (e.g., a protected left turn) is defined as a unique phase with a corresponding index in SUMO rather than a sub-phase.

\textbf{Reward functions: } The reward is an integral part of reinforcement learning and should represent the desired collective goal of the process while being local to each agent. It is calculated after a minimum fixed delay, $\Delta t$, following an action since the outcome of the action will not immediately be reflected in the traffic state. This delay is set to be the same as the minimum green time. Let $r_i^t$ denote the reward received by agent $i$ at time $t$. There are four well-known reward functions provided in the sumo-rl library and we test each one of them separately: 
\begin{itemize}

    \item \textit{Difference of wait times} calculated as the sum of the waiting time of all vehicles in all incoming lanes in an intersection minus the same sum from the previous time the reward function was called.
    $$r_i^t=\sum_{\ell \in \mathcal{L}_i^{\text {in }}} W_{\ell}^t-\sum_{\ell \in \mathcal{L}_i^{\text {in }}} W_{\ell}^{t-\Delta t}$$
    where $W^t_\ell$ is the total waiting time of vehicles on lane $\ell$ at time $t$.

    \item \textit{Average speed} is the time mean speed of all vehicles in incoming and outgoing lanes of the intersection normalized over the maximum allowed speed.
    $$r_i^t=\frac{1}{\left|\mathcal{U}_i^t\right|} \sum_{u \in \mathcal{U}_i^t} \frac{v_u^t}{v_{\max }}$$
    where $v^t_u$ is the speed of vehicle $u$, $v_\text{max}$ is the maximum allowed speed, and $\mathcal{U}_i^t$ is the set of vehicles in the vicinity of intersection $i$.

    \item \textit{Queue length} reward is the negative of the total vehicles queued at the intersection.
    $$r_i^t=-\sum_{\ell \in \mathcal{L}_i^{\text {in }}} n_{\ell, \text { stopped }}^t$$

    \item \textit{Pressure} reward is the difference between the number of vehicles leaving and approaching the intersection. 
    $$r_i^t=\sum_{\ell \in \mathcal{L}_i^{\text {out }}} n_{\ell}^t-\sum_{\ell \in \mathcal{L}_i^{\text {in }}} n_{\ell}^t$$
    where  $\mathcal{L}_i^{\text {in}}, \mathcal{L}_i^{\text {out }}$ are the sets of incoming and outgoing lanes at intersection $i$, $n_{\ell}^t$ and $n_{\ell, \text { stopped }}^t$ are the total and stopped vehicle counts, respectively.
    
\end{itemize}

The training was performed based on predefined vehicle flows specifically designed to cause congestion in the downtown region of the map where intersections are arranged in a grid. Vehicles moving into and out of the central area give the Q-learning agents a diversity of samples to learn from, resulting in a more comprehensive Q-table to choose actions during deployment. Finally, the list of traffic light agents is saved to a Python pickle file.

\subsection{Reinforcement Learning Agent Deployment in SUMO}
The CARLA-SUMO co-simulation takes a running instance of CARLA and starts a SUMO instance with the same road network, subsequently pairing the two using a bridge helper module. A single time step (tick) of this simulation is 0.05 seconds during which the SUMO simulation updates all vehicle locations and synchronizes with CARLA. We import the traffic light agents from the previously saved pickle file during the setup stage of the simulation. Since the same road network is used, the traffic light IDs in SUMO are the same as the ones used in training, and the bridge helper module is used to find the corresponding intersections in CARLA. 

After every 100 ticks, equivalent to 5 seconds, each traffic light has its state updated using the information from the camera detections and the act function is called. This function looks up the state in the agent's Q-table and outputs the index of the next green phase to be selected for the intersection according to the maximum Q-value. If the new phase is to be different from the old one, the lights are set to the yellow phase that follows the current green phase for a fixed yellow duration of 3 seconds and then switched to the desired green phase. 

\subsection{Synchronization and Component Integration}

In a traffic co-simulation framework involving CARLA and SUMO with AI-based computation modules (computer vision and reinforcement learning), synchronization of data flows and components is critical to ensure seamless integration and accurate simulation results. The synchronization process ensures that the simulated environment and traffic signals are consistent across both platforms. The following steps outline the design and consideration to achieve this synchronization in our framework.

\subsubsection{Time Synchronization}
The simultaneous simulation in two different software platforms necessitates accurate and frequent synchronization, without which the representation of the vehicle states between the two simulators may diverge. The co-simulation takes care of updating the position of the vehicles in CARLA after they have been advanced in SUMO every 0.05 seconds. This includes spawning and removing vehicles that are starting or ending their routes. Since we do not create any vehicles in CARLA, the reverse process of updating to SUMO is not utilized. During the computer vision computation, all the images from the cameras are processed in one go and this pauses the simulation noticeably. However, the pause does not affect the functionality of the co-simulation and the data is synchronized as usual after the states are updated for the traffic light agents. 

\subsubsection{Coordinate System Alignment}
CARLA and SUMO use different coordinate systems, which requires precise mapping between the two. The bridge helper module translates coordinates from SUMO to CARLA and vice versa, ensuring that the positions of vehicles, traffic lights, and other objects are accurately represented in both simulations. This alignment is crucial for placing virtual cameras correctly and ensuring accurate vehicle detection.

The module is used to access traffic light pole locations in the CARLA environment, spawning virtual cameras at these locations on a predefined height above the existing light pole. The IDs and coordinates of the stopping line of each lane are matched between CARLA and SUMO to give the appropriate label to the deployed camera. The camera yaw angle is calculated from the vector between the light pole and the stopping line to ensure it is properly aimed. This automated setup is a key technical contribution that makes scaling to larger networks feasible, a step beyond single-intersection or manually configured testbeds.

\subsubsection{Camera Data as Real-Time Input to Traffic Light Agents}
The virtual cameras were configured to capture images at a frequency of once per second. The images are then processed using the YOLOv5 object detection algorithm to identify vehicles in the oncoming lanes. During this computation the simulation does not advance and the data is synchronized to the simulation once all states are updated. 

The MARL agents in our framework, trained using the sumo-rl library, require a structured state vector comprising features such as lane occupancy and queue length, typically derived from ground-truth simulator data. In contrast, the CARLA environment provides only raw image data from virtual cameras. Bridging this modality gap posed a key integration challenge. To address it, we developed a custom vision-to-state translation module that processes images from multiple intersection-mounted cameras using a YOLO-based object detector. The module filters detections by vehicle type, restricts analysis to a defined region of interest, and aggregates counts across cameras serving the same intersection. These counts are then mapped to the appropriate SUMO lane indices and transformed into occupancy and queue metrics compatible with the pre-trained agents' Q-table. This translation layer is essential to enabling reinforcement learning control based on visual perception and represents a core technical contribution of our system.

\begin{figure*}[!htb]
    \centering
    \includegraphics[width=0.95\linewidth]{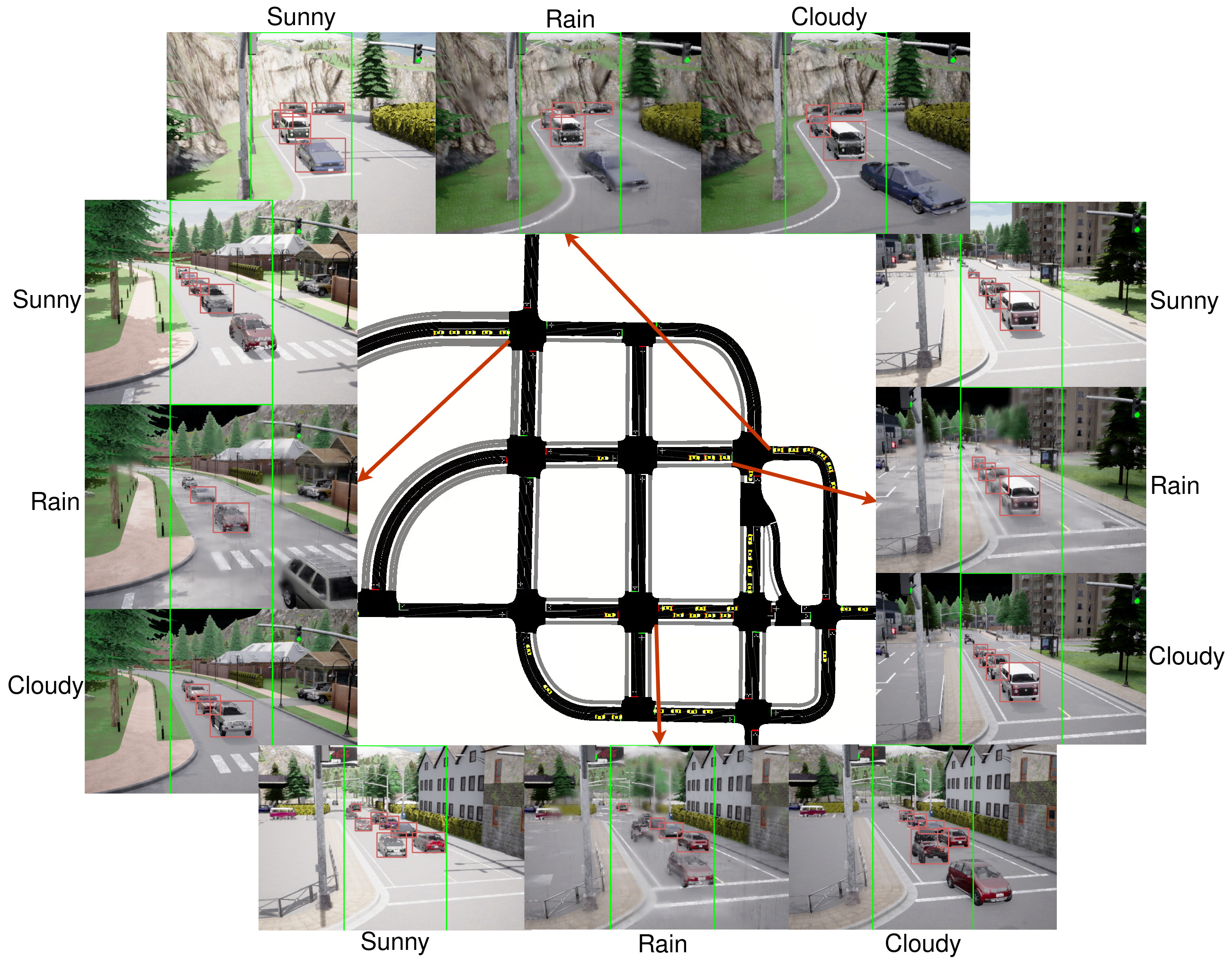}
    \caption{Camera views from multiple intersections in different weather and lighting conditions. Only the downtown part of the map is shown here. }
    \label{fig:weather}
\end{figure*}

The traffic light poles in CARLA are not organized in the same order as the states required for SUMO agents. Therefore, to identify the lane index that each camera is monitoring, the stop location of each CARLA traffic light is matched to the coordinates of the end of the corresponding lane in the SUMO intersection. Some views from the cameras in different conditions are shown in Figure \ref{fig:weather}.

\subsubsection{Action Execution and Feedback Loop}
Once an agent determines the next phase, the action is executed in the SUMO simulation. The CARLA simulation is updated to reflect the new traffic light phase, ensuring visual consistency. This feedback loop, where actions from reinforcement learning agents influence the simulation environment, and real-time data updates the agents' states, forms the core of the synchronized co-simulation framework.

The co-simulation operates at 20 Hz to maintain smooth vehicle dynamics, but real-time object detection on 43 cameras is computationally infeasible. To address this, we implemented a control loop that periodically pauses the simulation (e.g., every 5 seconds) to perform batched detection across all cameras. Upon resuming, the system ensures data consistency, allowing agents to act on a coherent snapshot of the environment. This asynchronous coordination is essential for practical vision-in-the-loop control.

\subsubsection{Error Handling and Robustness}
To handle potential errors and ensure robustness, the framework includes mechanisms for error detection and correction. For example, if a camera fails to capture an image, the last known state is used to update the reinforcement learning agent. If a state is not found in an agent's Q-table, no action is taken, but if inaction leads to a phase being stuck for over 60 seconds, phase change is forced. Additionally, the co-simulation synchronization module logs discrepancies and attempts to resolve them in real-time to maintain the integrity of the simulation.

\section{Experimental Analysis}
\subsection{Platform and Parameter Settings}
The experiments were executed on a desktop computer with 13th Gen Intel® Core™ i9-13900K CPU, 64 GB RAM, and NVIDIA RTX 4090 graphics card. The road network chosen for traffic signal optimization was ``Town04'', a map included in the CARLA software. It contains a downtown grid-like region surrounded by a highway. The East-West signal spacing is about 40-50 meters, while the North-South spacing is 50-70 meters. Entry and exit from the downtown are mediated by traffic lights, and the whole map contains 12 traffic lights in total. As part of the RL process, each traffic light is a Q-learning agent with access to its local state observation of the current phase, minimum green time indication, and the occupancy and queue for each incoming lane. The traffic signals operate independently: they are uncoordinated, possess no preprogrammed offsets, and do not exchange information with one another. The 9 downtown signals share the same cycle length in the case of static timing experiments. We manually specify traffic flow demands to cause congestion in the downtown grid and ample activity into and out of that region. These flows are defined as origin and destination, so the routes are not fixed and the vehicles dynamically determine their path from estimated travel time within the simulation. With static signals, one run of simulation lasts 2770 seconds in simulation time. 

\begin{figure}[!htb]
    \centering
   \subfloat[Average system speed during training \label{fig:RLspeed}]{%
        \includegraphics[width=0.75\linewidth]{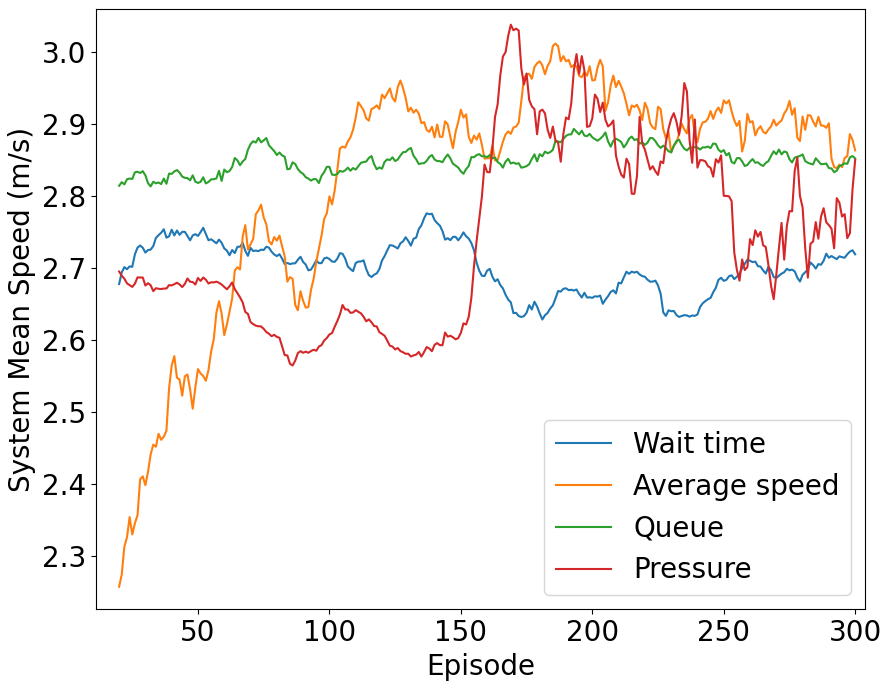}}  

\subfloat[Average wait times at one intersection \label{fig:RLwaittime}]{%
        \includegraphics[width=0.75\linewidth]{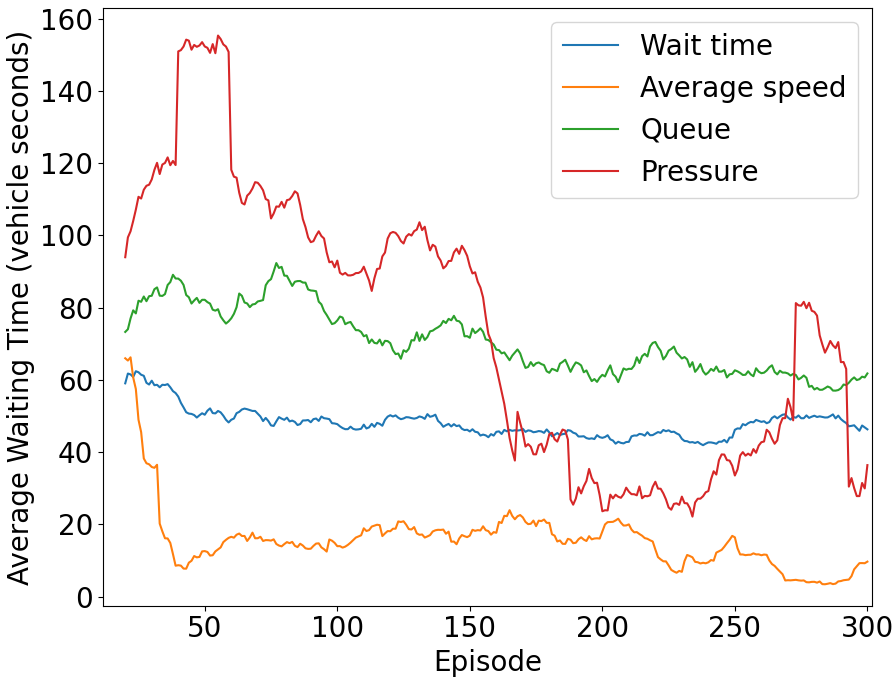}}

    \caption{Reinforcement learning using four different reward functions for 300 episodes}
\end{figure}

The learning progress of the agents is shown in Figure \ref{fig:RLspeed} which shows the average speed of vehicles in the network over the training run. Agents with average speed and pressure based reward showed significant network-wide improvement during training. Figure \ref{fig:RLwaittime} depicts the average wait times at one chosen intersection which also shows major variation for pressure and average speed based reward, and only minor improvement for queue based reward. The process resulted in four sets of trained agents, each set corresponding to a reward type, which were then saved as Python pickle files to be later used for evaluation in the co-simulation.

\subsubsection{Hyperparameter Tuning}
To optimize the training process for the various MARL agents across different reward functions and environmental conditions, a systematic hyperparameter tuning approach was employed. Specifically, the learning rate ($\alpha$) was incrementally adjusted from 0.1 to 0.001, the discount factor ($\gamma$) was varied within the range of 0.5 to 0.99, and the initial exploration rate ($\varepsilon$) was tuned from 0 to 0.05. A consistent set of hyperparameters was maintained for all experiments, encompassing different reward functions and traffic flow scenarios, to ensure comparability of results. The number of training episodes was also varied up to 5000. Having found satisfactory values that give repeatable results, the multi-agent RL was ultimately run with each reward function for 300 episodes with $\alpha = 0.0071, \ \gamma = 0.97$ and an initial $\varepsilon$ value of 0.05 which decays to 0.005.

\subsubsection{Co-simulation Testing Parameters}
The synchronization file for CARLA-SUMO co-simulation was modified to carry out the experiment. A camera was spawned on each traffic light pole resulting in 43 total cameras, and the corresponding lane was matched using the stop location coordinates. Each camera updates an image in the form of a numpy array which is then used as input to a pytorch instance of YOLOv5 to detect vehicles within $240 \times 480$ rectangle in the center of the frame. If the confidence score of the detection is higher than 0.5, it is counted as a vehicle belonging to the respective lane. Another function collects the required information and calculates occupancy and queue length for each intersection to form the state vector to be used as input for the trained agents. If the agent Q-table does not contain the state, the action is not updated in the simulation and its phase does not change. Otherwise, the phase is set to the recommended one after going through the appropriate yellow phase. 

There were two traffic demand scenarios tested in the co-simulation: medium and heavy. Both had the same routes but the heavy demand had higher vehicles per hour flow (591 vehicles at 200 to 400 veh/hr) for all routes as compared to the medium demand (387 vehicles at 100 to 300 veh/hr). The origin and destination for the demands were the same as the ones used for the MARL but differed in start and end times to ensure that we were not simply testing on the training scenario while also avoiding radically different traffic conditions. Congestion was observed in all scenarios, the last car entered at 1100 seconds into the simulations and the medium demand scenario lasted for 1700 seconds while heavy demand scenario ran for over 2800 seconds. The most congested signal in the medium demand scenario experienced an average West-East demand ($v$) of 700 veh/hr while having a theoretical capacity ($C$) of 1800 veh/hr, and with a static green time ($g$) of 22 seconds out of a cycle length ($T$) of 50 seconds. The $v/C$ ratio can be calculated as  
$$\frac{v}{C} = \frac{v}{s \times g / T} = 0.88$$
which is a level of service (LOS) `E', meaning unstable flow with high delays. Consistent queue spillback was observed for this intersection in the baseline simulation. Similarly, for the heavy demand scenario the same intersection received a demand of 1000 veh/hr, resulting in a $v/C$ ratio of 1.26, which is a considerably oversaturated LOS `F'.

\subsection{Co-Simulation Results}

\subsubsection{Traffic Signal Control with MARL}

Baseline simulation with static timing traffic lights was carried out for the two demands and it showed an average vehicle speed of 6.11 m/s and a total travel time of 145,144 seconds in the medium traffic, and 4.57 m/s average speed with 355,699 seconds total travel time in the heavy traffic. Table \ref{table1} shows the performance of the trained RL agents controlling the traffic lights using counts from the vehicle detections to form the observation vector. Out of the four variants tested, the agents trained with the reward function based on the average speed of vehicles near the intersection performed the best in both test scenarios by every measure. It is important to mention that the rewards are not available at test time, only the observation of lane occupancy and queue length is provided based on the camera sensing. 

The average waiting time is the duration of time a vehicle spends stopped, while average time lost is derived from the extra time the vehicle spent on a road by driving under the speed limit due to traffic, so the lost time includes the waiting time. In the heavy traffic scenario the agents trained on wait time rewards improve the wait times over the static traffic lights but still cause an increased time lost, indicating that the system prioritized stopped vehicles but overlooked the significant impact that frequent starting and stopping has on reducing traffic speed and throughput. 

Table \ref{table2} shows the results of the same simulations but with ground truth data of the lanes available to each agent. This includes an adaptive actuated traffic light controller that works based on the time-gap between vehicles. It operates by prolonging the green phase for consecutive traffic, with a maximum duration of 60 seconds. In both traffic scenarios the best trained agents were able to outperform the actuated signals, except in the metric of average speed during heavy traffic. This can be attributed to the actuated signal's tendency to prioritize overall vehicle movement by extending green phases for major thoroughfares, which can lead to a higher average speed across the network. However, this often comes at the cost of longer waiting times for vehicles on minor flows, as the system might not be as effective at minimizing individual vehicle delays at intersections. In contrast, the MARL agent, even when trained with an average speed reward, appears to learn a more sophisticated policy that balances network throughput with minimizing individual vehicle delays. For heavy traffic, minimizing stopped time (waiting time) is paramount for preventing gridlock and improving overall network efficiency, even if it means slightly lower peak speeds. This highlights a fundamental trade-off in how different adaptive control strategies optimize traffic, and suggests that reinforcement learning can offer distinct advantages in prioritizing local congestion relief and driver waiting times in extreme congestion.

It can be seen from Table \ref{table2} that the correct knowledge of the state of the intersection results in a better performance as compared to the same agents in Table \ref{table1} except in one case. Wait time and pressure reward based agents take a significant performance hit when working with noisy sensing, while the effect on the agent trained with average speed reward performs less than 10\% worse as compared to the ground truth scenario with medium traffic. The difference is more pronounced in heavy traffic. The relative robustness of agents with average speed reward indicates that while the other agents are learning first order relationships, speed based reward provides temporal insight to the Q-table. Since wait time and pressure are both directly linked to queue length and occupancy, agents with perfect information develop an overfitted estimate of the Q-value of taking a certain action, whereas imperfect observations during testing cause a small difference in the state vector from one measurement to the next, resulting in a significantly different Q-value. On the other hand, being rewarded with average speed, an agent will prioritize staying in the same phase for longer if the state does not vary too much between observations.

\begin{table*}[!htpb]
\caption{Simulation performance measures of statically configured traffic lights as compared to agents trained using different reward functions using computer vision (YOLOv5) for observation}
\renewcommand{\arraystretch}{1.25}
        \centering
    \begin{tabular}{lccccc}

                & Static &  Wait time  & Average speed & Queue & Pressure \\ \hline \hline
       \textbf{Medium traffic} &     &  &  &  &   \\
       Mean speed (m/s) & 6.11  & 7.68  & \textbf{9.13}   & 5.66   & 6.57    \\ 
   Total travel time (s) & 145,144  & 103,374 & \textbf{83,151} & 163,975 & 151,293  \\
       Mean waiting time (s)  & 137  & 39.9 & \textbf{28.7}   & 111   & 108         \\
       Mean time lost (s) & 228  & 121 & \textbf{68.4}   & 276   & 244            \\
       Mean trip duration (s) & 375  & 267  & \textbf{215}  & 424  &  391            \\ \hline
       \textbf{Heavy traffic} &     &  &  &  &   \\  
        Mean speed (m/s) & 4.57 & 4.21 & \textbf{6.02}   & 4.62    & 3.78   \\ 
   Total travel time (s) & 355,699  & 378,305 & \textbf{208,443} & 323,366 & 451,718 \\
       Mean waiting time (s) & 321  & 286  & \textbf{77.9}   & 184       & 287     \\
       Mean time lost (s) & 458  & 497  & \textbf{212}    & 403     &   621      \\
       Mean trip duration (s) & 601 & 640 & \textbf{354}    & 547      &  764    \\ \hline
    \end{tabular} \\

    \label{table1}
\end{table*}

\begin{table*}[!htpb]
    \caption{Simulation performance using actuated lights and trained agents agents trained using different reward functions with ground truth observations }
\renewcommand{\arraystretch}{1.25}
    \centering
    \begin{tabular}{lccccc}
                &    Actuated &   Wait time  & Average speed & Queue & Pressure \\ \hline \hline
       \textbf{Medium traffic}  &  &  &  &    &   \\
       Mean speed (m/s) & 9.37 &  9.48    & \textbf{9.84}  &  4.69  &   9.39  \\ 
   Total travel time (s) & 85,273  & 79,451  & \textbf{76,316} & 243,401  & 80,846  \\
       Mean waiting time (s) & 23.48 & 21.1  &   \textbf{19.4} &  225  &   20.3      \\
       Mean time lost (s) & 73.3 &   58.7 & \textbf{50.6} &  481  &     62.3   \\
       Mean trip duration (s) & 220 & 205 &  \textbf{197} &  629 &       209       \\ \hline
       \textbf{Heavy traffic}   &  &    &    &     &   \\  
        Mean speed (m/s) & \textbf{8.03}  & 7.00  & 7.11  &  3.67   &   5.68  \\ 
   Total travel time (s) & 181,715  &  163,782  & \textbf{158,245} & 559,992 &  227,577  \\
       Mean waiting time (s) & 92.96 &  42.3  &  \textbf{36.3}  &   400  &  108    \\
       Mean time lost (s) & 163 &   136  &  \textbf{127} &   804  &    243     \\
       Mean trip duration (s)  & 308 &  277  &  \textbf{268} &      947  &   385   \\ \hline
    \end{tabular} \\
    \label{table2}
\end{table*}

Queue length based reward proved to be the worst performer in both cases but the drastic difference with ground truth information is definitely anomalous as the camera detection version had better metrics. A closer look at the simulation revealed that during congestion the traffic backs up onto the highway, which is not part of the short lane being observed by the traffic light. Therefore, the agent always observes that the queue is full, and any action it takes will not change the size of the queue at the next measurement. As a consequence, that particular agent never learns the real value of giving the green phase to the incoming traffic. Interestingly, the reason that the camera detection version performs better is that due to a failure to detect all the cars, or detecting extra cars in the adjacent lane, the count deviates from the ground truth causing different rows of the Q-table to be accessed, forcing the agent to change the phase.

\subsubsection{Video Detection}
Cameras were deployed on top of traffic light poles facing the lanes they service to monitor and count vehicles using YOLOv5 object detection. This setup aims to simulate real-world deployment utilizing minimum resources and capturing the state of the lanes only once every 5 seconds when it is time to make a decision about the signal phase. A limited region of 240 pixels wide in the middle of the image frame was used to count vehicles, so the parked cars seen in the first camera image of Figure \ref{fig:framework}, for example, were not counted as observed vehicles.

\begin{figure}[!htb]
    \centering
    \subfloat[YOLOv5 histogram of errors \label{fig:yolo5hist}]{
        \includegraphics[width=0.75\linewidth]{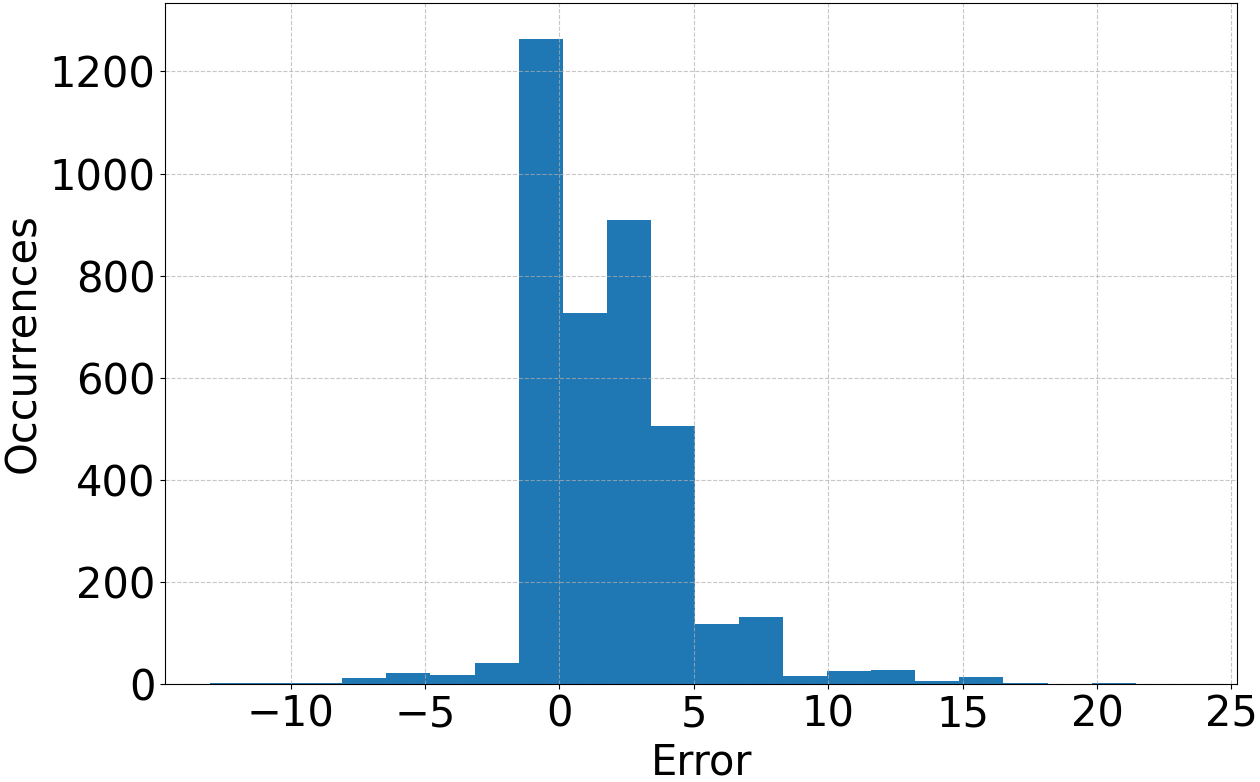}}
        
    \subfloat[YOLOv8 histogram of errors \label{fig:yolo8hist}]{
        \includegraphics[width=0.75\linewidth]{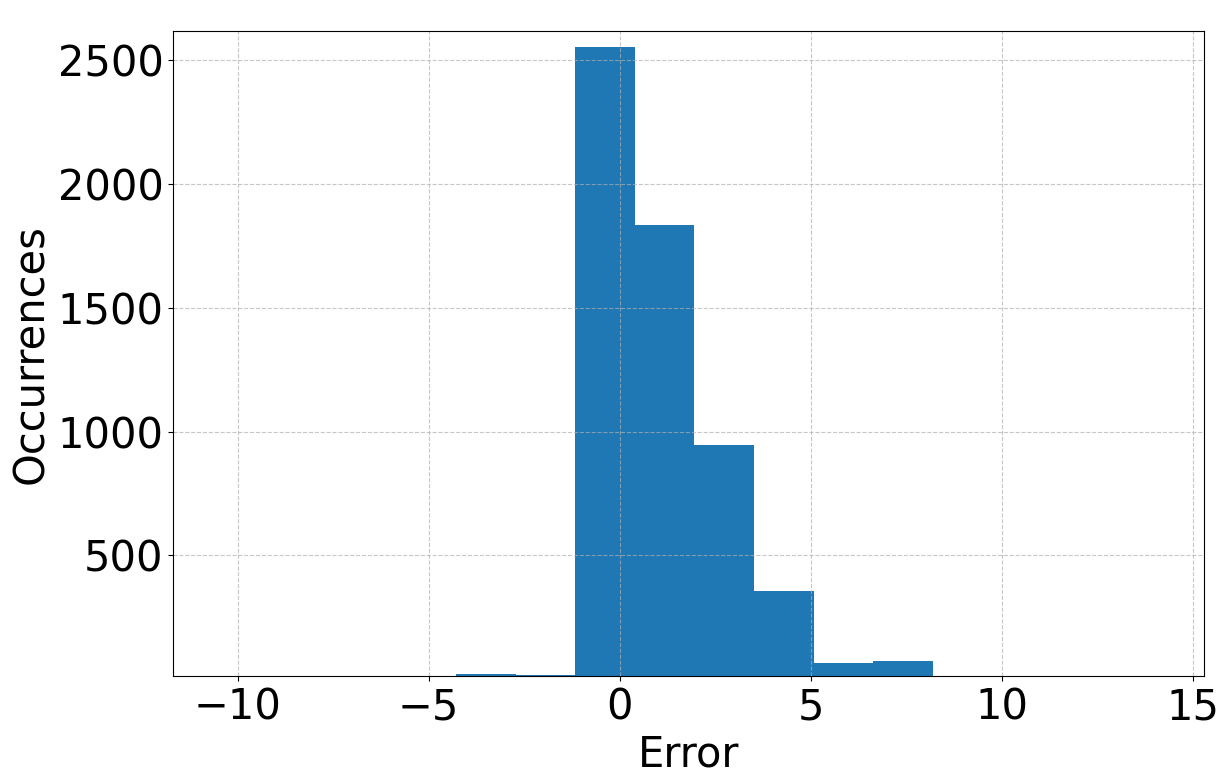}}
        
    \caption{Computer vision based counting errors using different YOLO versions over one simulation run}
\end{figure}

The lack of calibration and fine-tuning led to some inaccuracies in vehicle counting. In some instances, the system under-counted the number of cars present at the intersection, especially cars that were far away or partially occluded from view. This is also compounded by the fact that a few of the approaches to intersections are not straight, and the lane curves out of view of the camera. More frequently, however, the cameras over-counted by including cars that were moving away from the traffic signal. A graph of the distribution of errors in the network during heavy traffic simulation is presented in Figure \ref{fig:yolo5hist}, showing in some instances over-counting up to 20 extra cars, and in others missing 10 cars in the whole network, but over-counting was more common. Overall, in a network of 12 intersections comprising of 43 approaches the mean absolute error was only 2.21 cars while the root mean square of errors was 3.48 cars. 

YOLOv8 was also tested in place of YOLOv5 as it is a newer and more powerful version of the model. The particular pretrained checkpoint used was YOLOv8m with 25.9 million parameters. The accuracy was higher with a mean absolute error of 1.73 and root mean square error of 2.80. The distribution plot of errors is shown in Figure \ref{fig:yolo8hist} which still shows a positive skew, but the errors are much lower both in terms of extremes and on average.
However, YOLOv8 was much slower to execute with the simulation taking twice as long as real-time. The simulation was run with heavy traffic using the same trained MARL agents as before. Table \ref{table3} shows that the increased accuracy did result in improved system performance for all agents except those with queue reward. For the best performing agents, the average speed increased from 6.02 m/s to 6.78 m/s, which is much closer to the 7.11 achieved in the ground truth case. Total travel time decreased significantly from 208,443 seconds to 171,526 seconds and the average waiting time came down from 77.9 seconds to 41.4 seconds. In all cases, the performance of the agents using YOLOv8 was closer to the performance from agents that used ground truth observations.

\begin{table*}[!htb]
\caption{Simulation performance using YOLOv8 in heavy traffic scenario}
\renewcommand{\arraystretch}{1.25}
    \centering
    \begin{tabular}{lcccc}
       \textbf{Heavy traffic} & Wait time  & Average speed & Queue & Pressure  \\  \hline \hline
        Average speed (m/s) & 4.75 & \textbf{6.78}  & 4.52    &  4.43 \\ 
   Total travel time (s) &  315,682 & \textbf{171,526} & 360,210 &  391,806 \\ 
       Average waiting time (s) & 190 & \textbf{41.4}   &  230     & 270 \\ 
       Average time lost (s) & 390   & \textbf{150 }    & 466      & 521  \\  
       Average trip duration (s) & 534 & \textbf{290}   &  610    & 663 \\ \hline
    \end{tabular} \\
    \label{table3}
\end{table*}

\begin{figure}[!htb]
    \centering
    \includegraphics[width=0.95\linewidth]{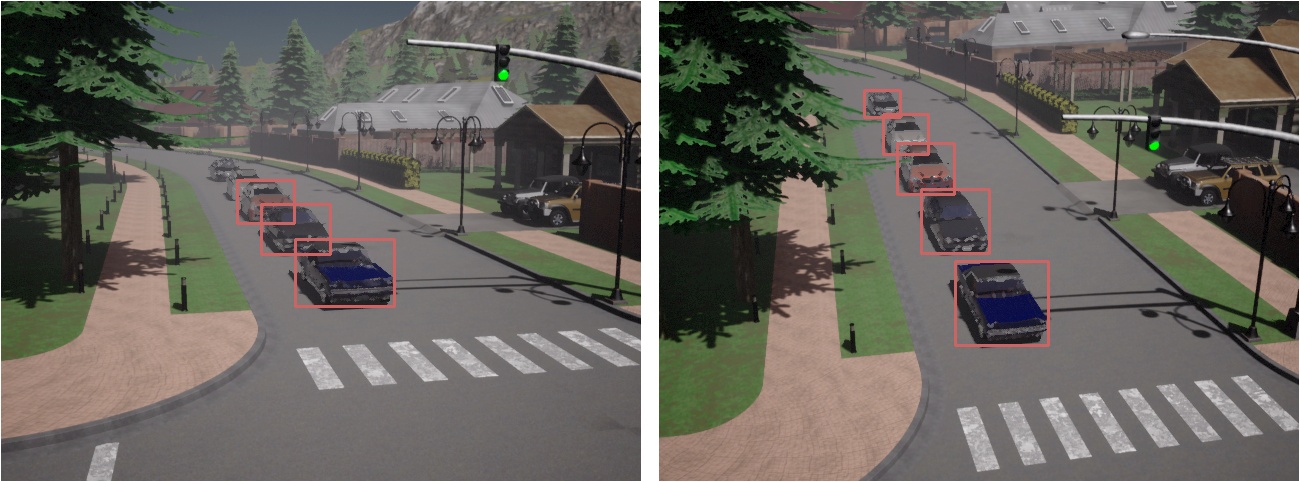}
    \caption{Camera views from the top of traffic light pole compared with much higher mounting }
    \label{fig:angle}
\end{figure}

The vehicle detection method can be fine-tuned for this specific purpose by generating training data in CARLA to train a better detector. Additionally, there are more sophisticated computer vision solutions available, for example YOLOv9 \citep{wang2024yolov9} and Detection Transformer \citep{ouyangzhang2022nmsstrikes}. Our choice was based on ease of use and fast processing within the co-simulation as there are 43 cameras. The angle and cropping of each camera can also be manually adjusted to focus only on the required region of the image. For practical realism, the location was chosen as the top of the traffic light pole, however the camera can be mounted much higher to give a better vantage. An example of this is shown in Figure~\ref{fig:angle}, where the left image is from a camera at 5 meters above the ground and the right one is from 10 meters, which is able to successfully detect two additional cars.

A tracking algorithm like DeepSORT \citep{wojkeDeepsort} can be integrated with detection to boost accuracy since tracking holds across frames even if detection is momentarily lost. This would require more frequent sampling, at least 5 frames per second. In the real-world implementation this would be feasible as in such distributed control systems edge computing may be used to process the data at the intersection, but in a simulation environment, performance matters. Without computer vision the simulation runs in real-time but during heavy traffic 100 seconds of simulation time corresponded to 145 seconds of wall clock time with YOLOv5 and just over 200 seconds with YOLOv8 due to the combined computation load of vehicle detection from all cameras.

\subsection{ Analysis of Learned Policies and Failure Modes}
\subsubsection{ Interpretability and Learned Signal Patterns}
The Q-learning algorithm employed in this study inherently supports a degree of interpretability through its Q-table structure. Each Q-table is a matrix that maps specific environmental states to the expected cumulative reward for taking a particular action from that state. The state of an agent is defined by the current phase index, a boolean indicating minimum green time elapsed, and vectors for occupancy and queue length in incoming lanes. The action space corresponds to the index of the next green phase to be activated.
By inspecting the Q-table for a given intersection, one can observe the learned policy directly: for any given combination of phase, minimum green time status, and observed lane occupancy and queue lengths, the table indicates the action (next phase) that the agent has learned to be optimal. For example, an entry in the Q-table looks like this: 
$$\text{State: } (0, 1, 0, 7, 0, 9, 0, 7, 0, 0) \rightarrow \text{Q-values: } (51.12, 0.01)
$$
Meaning that the signal is in phase 0, corresponding to a green light for East-West traffic, the minimum green time has elapsed, and there are 7 cars coming from the West which are all stopped to form a queue, while there are 9 cars coming in from the East which are all moving as the queue value is zero. In this scenario the Q-value is determined to be high for remaining in phase zero.

This direct mapping allows for a transparent understanding of the agent's decision-making process under various traffic conditions. For instance, if the Q-table consistently shows a high Q-value for extending a green phase when a high occupancy and queue are detected on that approach, it indicates the agent has learned to prioritize clearing that congestion. Conversely, a low Q-value for a phase might indicate the agent has learned to avoid it under certain conditions. This explicit, tabular representation of the policy contributes to the regulatability of the control policy, as changes in signal assignment can be traced back to specific changes in the observed traffic state, similar to how traditional actuated systems operate based on predefined logic. The observed robustness of agents trained with the `average speed' reward function further suggests that these agents develop more stable and predictable policies, as they are less sensitive to instantaneous noise in observations compared to agents relying on more volatile metrics like waiting time or pressure.

\subsubsection{Failure Modes}
\textbf{Phase Lock-in:} Phase lock-in refers to a scenario where a traffic signal becomes stuck in a particular phase, preventing other movements from receiving a green light, leading to severe congestion. The implemented framework includes a specific error handling mechanism to mitigate this risk. As mentioned earlier, if a state is not found in an agent's Q-table, no action is taken, but if inaction leads to a phase being stuck for over 60 seconds, phase change is forced. This rule acts as a safeguard, ensuring that even if the MARL agent fails to provide a new action or gets into an unforeseen state, the system will revert to a basic, time-based control to prevent a complete gridlock at the intersection. This mechanism directly addresses the potential for phase lock-in by enforcing a maximum green time, a common practice in traditional traffic signal control.

\textbf{Queue Spillback:} Queue spillback occurs when a queue of vehicles extends from one intersection back into an upstream intersection, blocking traffic flow and exacerbating congestion across the network. The study observed an instance of this failure mode with the `queue length based reward' agent. Specifically, during heavy congestion, traffic backed up onto the highway, which was outside the observation range of the traffic light's camera. Consequently, the agent consistently observed a `full queue' state, and any action it took did not appear to change the queue size in its immediate observation. This led to the agent failing to learn the effective value of giving a green phase to the incoming traffic, as its reward signal was not sufficiently responsive to the actual queue dynamics extending beyond its local sensing. This highlights a critical limitation of purely local observation and reward functions in preventing network-wide congestion phenomena like spillback. While the current framework demonstrated this failure mode, future work can explore strategies to address it, such as incorporating a broader observation area, using network-level reward signals, or implementing explicit queue management strategies within the MARL framework, as some existing RL approaches for ramp metering have shown success in preventing queue spillback by estimating lower bounds for metering rates and replacing actions that violate these constraints.

\subsection{Discussion: Extension of Co-simulation Framework to Digital Twin}

Co-simulation is a useful, and perhaps necessary, step before real-world testing for transportation research. It can enable thorough examination of automation processes that involve sensing and control in a variety of scenarios that are likely to be faced after deployment. Artificial intelligence powered applications including traffic signal control, variable speed limits \citep{zhang2024field}, parking management, etc. all rely on sensing and control feedback loops integrated into the diverse urban environment. There can be many variables that affect the system and noisy sensing is one example. Robustness to those conditions should be proven in co-simulation environments for practical feasibility. The framework developed in this paper lays the foundation for further extension into the domain of digital twin technologies. A digital twin for urban traffic systems involves the real-time integration of physical infrastructure and simulation environments, enabling continuous monitoring, simulation, and optimization of the entire traffic network. Through edge computing, real-world sensors like infrastructure cameras, connected autonomous vehicles (CAVs), and even roadside units can provide continuous streams of traffic data \citep{azfar2023incorporating}. This real-time data can be integrated with the co-simulation environment, enabling online learning and adaptive control strategies.

Incorporating real-world data into the CARLA-SUMO co-simulation via edge computing offers numerous advantages, particularly when it comes to handling noisy and incomplete sensor data. In real-world scenarios, environmental conditions, hardware limitations, and communication delays can all degrade the quality of sensor input. By simulating these real-world imperfections in the co-simulation environment, more robust control algorithms can be trained that are resilient to noisy data and infrastructure variability. For instance, the use of camera-based vehicle detection can be tested not only under ideal conditions but also under challenging scenarios such as low visibility, occlusions, and adverse weather conditions. Additionally, the 3D digital twin of the specific location can provide the precise learning environment for the MARL agents to be deployed. 

A key feature of the digital twin system is the possibility of continuous learning using real-time data. While simulated traffic based on estimated demands and orderly driving behaviors provides a good foundation for reinforcement learning, the real-world chaotic traffic, dynamic travel demands, and evolving urban landscape require online learning from the collected data. Real-world traffic data, when combined with synthetic data from the co-simulation, can be used to capture both short-term fluctuations in traffic patterns as well as long-term trends. Intuitively, this will enhance the quality of the MARL agent Q-tables as they will better reflect the advantage of taking a certain action in the physical road network. The scalability of the digital twin architecture ensures that the lessons learned from localized simulations can be extrapolated to broader contexts, making it a valuable tool for city planners and transportation authorities aiming to implement smart city initiatives.

Such digital twin frameworks are directly applicable to CAV deployment, as they enable evaluation of CAV-infrastructure interactions under realistic sensing conditions, facilitating safer and smoother transitions from simulation to field implementation. Furthermore, CAVs could play a pivotal role in enhancing the co-simulation by acting as mobile sensors. These vehicles are equipped with advanced sensor arrays, including LIDAR, radar, and cameras, allowing them to gather rich environmental data as they navigate the traffic network \citep{ke2023lightweight}. By leveraging cooperative sensing from CAVs, the system can implement distributed learning algorithms that optimize traffic signal timing and traffic flow in real-time. CAVs can also be used to detect and mitigate traffic bottlenecks by actively controlling traffic patterns, especially on highways, to reduce shockwaves and smooth out traffic congestion \citep{zheng2020smoothing}.

A complete digital twin co-simulation framework in the context of traffic signal control will consist of real-time data collection, simulation, and feedback loops for continuous optimization of traffic networks. Sensing infrastructure can be enhanced with cameras and CAV data in addition to loop detectors by aggregating and preprocessing on roadside edge computing units. The edge computer will host the vehicle detection module and compare and combine results with loop detectors, traffic signal cabinets, and connected vehicle data before transmitting the relevant information to the centralized digital twin. At the central server the data from all edge devices will be organized and used for online reinforcement learning. Meanwhile the co-simulation will simultaneously be performed on the digital twin providing synthetic variation to the collected data to further aid training. In addition to MARL for TSC, the digital twin platform could enable the development of short-term traffic forecasting algorithms \citep{VLAHOGIANNI20143}. 

Finally, to close the digital twin information loop, the system will send updates to edge devices based on learning outcomes. In the case of MARL these will be updated Q-tables, while CAVs can benefit from network wide data to adjust their individual goals. Digital twin predictions can be used to manage traffic on a larger scale through dynamic road messages for rerouting and speed limits, signal timing coordination, ramp meter adjustment, and emergency vehicle route planning. This digital twin infrastructure will enable intelligent transportation systems to optimize traffic flows, reduce congestion, and improve overall urban mobility.

\section{Conclusion}
Intelligent transportation systems rely on optimization in simulation before deployment to the real-world and face challenges associated with erroneous sensing that must be addressed beforehand. To this end, network-wide multiagent traffic signal control was performed using reinforcement learning and tested with computer vision based vehicle counting in a co-simulation environment. Multiagent reinforcement learning with different reward functions was tested on a SUMO road network, and then the learned agents were deployed in the CARLA-SUMO co-simulation using cameras mounted on traffic lights for vehicle counting as input to the agents. Vehicle detection was performed using YOLOv5 and YOLOv8 against the four different types of MARL agents. While detection accuracy does improve performance, it was found that even with faulty detections, MARL agents trained to reward average speed performed favorably but other methods such as queue length based reward were highly sensitive to the errors. 

Beyond advancing adaptive signal control, the findings underscore the potential of infrastructure assisted reinforcement learning to support automated vehicles. By showing that vision driven control remains effective with imperfect sensing, the study contributes a practical step toward scalable CAV deployment in real-world urban networks.

The framework described in this paper can be applied and extended to traffic signal control of real world locations and support more complex scenarios involving cooperative perception, connected autonomous vehicles, and digital twins with field collected data. 
This study lays the groundwork for a scalable co-simulation framework integrating vision-based sensing with MARL for traffic control. Future work will focus on enhancing the perception pipeline through post-processing techniques such as confidence-weighted filtering and multi-object tracking (e.g., DeepSORT) to improve detection accuracy and temporal consistency. Incorporating custom-trained object detectors optimized for the CARLA environment may further reduce domain mismatch. Additionally, exploring the interplay between noise characteristics, state encoding, and reward formulation, particularly under highly congested conditions, could yield new insights into designing more robust MARL agents. Real-time feasibility of these enhancements could be supported through edge computing or sensor hardware optimization.

\bibliographystyle{apacite}
\bibliography{interactapasample}

\end{document}